# Heat loss and internal dynamics of Venus from lithosphere strength

Javier Ruiz[1,*], Alberto Jiménez-Díaz[2], Isabel Egea-González[3], Ignacio Romeo[1], Jon F. Kirby[4], Pascal Audet[5]

[1] Departamento de Geodinámica, Estratigrafía y Paleontología; Facultad de Ciencias Geológicas, Universidad Complutense de Madrid, 28040 Madrid, Spain
[2] Departamento de Biología y Geología, Física y Química Inorgánica. ESCET, Universidad Rey Juan Carlos, 28933 Móstoles, Madrid, Spain
[3] Departamento de Física Aplicada. Escuela Superior de Ingeniería. Universidad de Cádiz, 11519 Puerto Real, Cádiz, Spain
[4] Geodesy & Earth Observation, DTU Space, Technical University of Denmark, 2800 Kongens Lyngby, Denmark
[5] Department of Earth and Environmental Sciences, University of Ottawa, Ottawa, Ontario, Canada

* Corresponding author. E-mail: jaruiz@ucm.es

## ABSTRACT

The absence of plate tectonics and the young surface age (0.3-1 billion years) of Venus have led to diverse geodynamic models for Venus. The energetics of the Venusian interior drive these models; however, the lack of direct constraints on surface heat flow hampers their quantitative assessment. Here we present a global heat flow model for Venus, as well as estimates of the total heat loss, obtained from an inversion of geophysical data, including lithospheric effective elastic thickness, crustal thickness, and radioactive heat production. Heat flow on Venus is lower and less geographically structured than on Earth, with an average of 31 mW $m^{-2}$, but with highs associated to rifts systems reaching values typical of active terrestrial areas. The obtained total heat loss is 11-17 TW, similar to estimates of the total radioactive heat production. Therefore, at present, Venus proportionally dissipates much less heat than Earth. Furthermore, the calculated crustal temperatures imply that crustal melting or eclogitization are not dominant processes in the Venusian crust.

## INTRODUCTION

Despite their similar size and mass, Earth and Venus have very different internal dynamics that reflect contrasting modes of heat loss. On Earth[1-3], plate tectonics drives heat loss through lithosphere recycling, with a substantial contribution from hydrothermal circulation through oceanic plates, and a minor contribution from mantle plume (i.e. hot spot) activity. Plate tectonics creates significant topography and surface age differences, as well as a structured surface heat flow distribution[3], with high and low heat flow corresponding to mid-ocean ridges and old continental cores, respectively. In comparison, the surface of Venus (see Fig. 1) is more homogeneous, has lower relief, and shows abundant evidence of effusive volcanism. Models of the internal dynamics sought to explain these observations invoke a variety of processes[4]: The surface lid could be mostly sluggish with limited lithosphere movement[5]. Heat loss could be dominated by plume and rifting activity, raising significant amounts of volcanism or plutonism[6], which could even make the lithosphere squishy and fragmented[7]. Venus could be operating in an "episodic mode" alternating stagnant lid and lithosphere wholesale recycling (including outbursts of subduction and/or magmatism)[8,9], or even as a mix of processes in space and/or time[10-13]. Furthermore, early giant collisions might have had a significant role in the duration of volcanism and resurfacing[14]. However, the lack of constraints on heat loss from direct surface heat flow measurements for Venus hampers a quantitative appraisal of these models.

Surface heat flow can be estimated using proxies for the thermal structure of the lithosphere derived from satellite data. On Venus, local and regional heat flow estimates were previously obtained from modeling of lithospheric flexure[15-19], depth to the brittle-ductile transition[20-22], and topographic and crustal relaxation[23,24]. These estimates provided local and regional constraints on heat flow. On the other hand, average surface heat flow values were proposed using mass/radius scaling based on terrestrial heat flow or radioactive heat production[25-27], which requires assuming (explicitly or implicitly) similar dynamics for both planets.

In this study, we use a map of the spatial variations of the effective elastic thickness of the lithosphere ($T_e$) to generate a global, 2º × 2º surface heat flow model for Venus from the relation between lithosphere strength and thermal state. The calculated heat flow values represent the thermal state when the long-wavelength relation between topography and gravity was established. Because the Venusian surface is young[28], our map is representative of the thermal state of Venus in recent times (see Discussion). From the $T_e$ map, we use appropriate values for the relevant parameters (see Methods) that permit obtaining an upper limit for the heat flow in each pixel of the map. Using these heat flow upper limits, we calculate upper limits for the total heat loss of Venus, which can be compared with geochemical information and Earth-based scaling. These estimates provide quantitative constraints for the appraisal of geodynamic models of Venus.

## RESULTS

### Effective elastic thickness map

We obtained an updated version of the global map of the effective elastic thickness of the lithosphere by Jiménez-Díaz et al.[29] (Fig. 2) by inverting the spectral coherence function between topography and Bouguer gravity anomaly using the fan wavelet transform[30] (see Methods).

Two other global mapping for $T_e$ were previously proposed. The global map by Anderson and Smrekar[31] was constructed from admittance method[32] and is a map of intervals (each interval spanning ten kilometers), not being therefore easy to obtain mean values or heat flow determinations from it. Moreover, a $T_e$ analysis of grids generated from harmonic expansions are likely to be more robust when using the coherence than the admittance (see Methods). The map by these authors presents in general lower $T_e$ values than our map, but the trends (not the absolute values) of high and low $T_e$ for most regions are roughly comparable. However, the admittance method used[32] has been considered to generate spectral leakages and biases[33].

The global 5º × 5º $T_e$ map by Audet[34] was constructed through the joint inversion of admittance and coherence using a spherical wavelet method, and obtained determinations for a fraction of the surface where that inversion gave reliable results with that procedure. The average $T_e$ in these regions is 82 km. As a comparison, the average value for the same regions in our map is 43 km (similar but slightly higher than our global average of 40.5 km), whereas general trends (not the absolute values) of high and low $T_e$ are also roughly similar in both maps.

Our results are therefore between those obtained by previous global maps. Thus,

whereas recognizing the limitations in the gravity field accuracy (see Methods), we assume our $T_e$ map as a representation of the lithosphere strength of Venus.

**Heat flow model**

Surface heat flows were derived from a procedure relating $T_e$ and the temperature-dependent lithospheric strength envelope (which determines the mechanical thickness of the lithosphere, $T_m$)[35,36] and constraints from a global crustal thickness model. The relation between $T_e$ and $T_m$ depends on the curvature of the elastic plate, and it must be $T_e \leq T_m$. If plate curvature is zero, then $T_e = T_m$. This case gives the highest heat flow for a given $T_e$. Because the wavelet methodology does not permit a direct calculation of plate curvatures, we assume $T_e = T_m$, which gives an upper limit for the derived heat flows. This is useful for constraining the thermal evolution of a planetary body[36]. For non-negligible plate curvatures $T_e < T_m$, and the surface heat flows would be lower than our estimates. Therefore, due to the absence of estimates of plate curvatures from the wavelet method, we cannot present a rigorous lower limit calculation for the heat flow.

We calculate the heat flow from $T_e$ for each grid pixel of our map (details on the calculation procedure and assumptions are given in Methods). Our model considers a basaltic crust for Venus[37], with an appropriate nominal density of 2900 kg $m^{-3}$, and thermal conductivity of 2 W $m^{-1}$ $K^{-1}$. The possibility of a more felsic composition for crustal plateaus has been suggested[38,39] but its effect on total heat loss calculations would be minimal (see Discussion). Our model also considers crustal heat sources, which increases the calculated heat flow for a given $T_e$ (ref. 36), consistent with upper limit calculations. We use 0.45 μW $m^{-3}$ as upper limit for the average radioactive heat production based on measurements from Venera and Vega landers[40] (see Methods). For the lithosphere mantle we use a thermal conductivity of 3 W $m^{-1}$ $K^{-1}$, a value appropriate for mantle rocks at temperatures of at least several hundred degrees Celsius[41-43]. We construct the heat flow model for a strain rate of $10^{-16}$ $s^{-1}$, typical of active terrestrial plate interiors and widely used for Venus[19,44,45].

The obtained 2º × 2º heat flow map (Fig. 3a) has minimum and maximum values of 17 and 178 mW $m^{-2}$ respectively, and an average value of 31.2 mW $m^{-2}$. The map shows large-scale areas with either high or low heat flow values: high values (>35 mW $m^{-2}$) are mainly found on much the regional volcanic plains, including Akhtamar, Tahmina, Atalanta, Guinevere, Sedna and Lavinia Planitiae; intermediate values (26-35 mW $m^{-2}$) characterize Niobe Planitia and the highlands of Aphrodite and Ishtar Terrae; low values (<26 mW $m^{-2}$) are found on the regional plains of Northern Nsomeka and Aino Planitiae, on Hinemoa Planitia in the centermost of the region known as the BAT anomaly area[46], and on Themis Regio, located in one of the vertices of that area. However, there are areas with high heat flow values on the rifts system demarcating, or exterior to, the outer boundary of the BAT area, ~70 mW $m^{-2}$ on Parga and Devana Chasmata, and until ~180 mW $m^{-2}$ on Dali Chasma (Figure 4a,b).

The uncertainties in heat flow estimates derived from the uncertainty in $T_e$ determination (maintaining other parameter equal in order to obtaining an upper limit for the heat flow in each grid pixel; see Methods) are usually small (Fig. 3b), with an average of 0.8 mW $m^{-2}$, although locally can reach 9 mW $m^{-2}$ for high heat flow values.

The ratio between maximum and minimum values in the map for Venus is around ten, and the geographical heat flow pattern is very different and less structured than that observed on Earth[1-3]. Thus, Venus lost heat more homogeneously than Earth, whose interior cooling pattern is a direct consequence of plate tectonics.

**Total heat loss**

We use our heat flow map to calculate a nominal upper limit for the total heat loss of Venus of 14.4 TW. Considering the uncertainty in heat flow estimates due to $T_e$ determination and crustal thicknesses between 20 and 25 km (see Methods), the total heat loss could be between 14.0 and 14.8 TW. Otherwise, if crustal heat sources are not considered (i.e., using linear thermal gradients), the total heat loss falls to 10.7-11.9 TW.

As a sensitivity test, we calculated the total heat loss for a faster strain rate of $10^{-15}$ $s^{-1}$, typical of terrestrial plate boundaries, obtaining 15.6 TW (average heat flow of 34.0 mW $m^{-2}$). Thus, considering the possibility of regionally high strain rates, uncertainty in heat flow estimates, and average crustal thickness between 20 and 25 km (see Methods), the total heat loss could even be 16.1 TW, although we favor a value close to our nominal value, because such a faster strain rate would be excessively high for most Venusian regions (see Methods).

The total heat loss obtained here is much lower than the terrestrial values scaled to Venus. Scaling the updated estimate for Earth of 40-42 TW (ref. 3) to Venus would give 33-34 TW. The terrestrial total heat loss is much higher than the radioactive heat production, which lies between 14 and 20 TW, depending on the geochemical model considered[47-50] (see Methods). Thus, Earth is losing two or three times more heat than radioactively produced, which implies vigorous interior cooling. If we scale the terrestrial radioactive heat production to Venus we obtain a range between 11 and 16 TW. Heat loss and radioactive heat production values are therefore comparable for Venus: the heat loss of this planet would be, at most, around 35-50% higher than its heat production but could be lower. This can be expressed in terms of the bulk Urey ratio, defined for a bulk planet as the ratio between the total radiogenic heat production and the total surface heat loss[51]. Whereas this value for the Earth is 0.3-0.5, we obtain a value ≥0.7 for Venus, implying that the bulk interior of this planet is moderately cooling down or even heating up.

## Temperature at the base of the crust

Under particular conditions, large-scale basalt melting[18], or eclogitization and associated delamination[52], can occur in the lower crust. This might affect the stability of the crustal structure and, therefore, the consistency and validity of our results. We calculate the temperature at the base of the crust from our nominal heat flow results and crustal thickness model (Fig. 5) and find temperatures exceeding the basalt solidus or liquidus (see Methods) only in rare cases (see Fig. 6). Therefore, our analysis does not predict large-scale melting of the crust. Since our heat flow calculations are upper limits, the actual crustal thermal profile may not reach the basalt solidus at these locations. Similarly, eclogitization is not expected for our temperature profiles (see Fig. 6). The basalt-eclogite transition could occur in locations with a crustal thickness greater than 40 km, but the expected temperature at the base of the local crust exceeds the temperature at which the eclogite transition takes place. However, because our heat flow values are upper limits, we cannot discard the possibility of eclogitization in regions with thicker crust, if actual heat flows and crustal temperatures in these regions are lower than derived from our model. Thus, our regional heat flow estimates (and hence the total heat loss of Venus) are consistent with crustal stability. If the average crust thickness is lower than that in our model, then the conditions for stability would be even easier to be satisfied.

## DISCUSSION

Our results reveal that the interior dynamics of Venus is very different from that of the Earth based on both global heat budget and heat flow geographical distribution. Indeed, we show that Venus loses proportionally less heat than Earth, with much lower relative variation across its surface. The estimated heat flows span a wide range of values including those obtained by previous works for specific features or terrains on Venus[15,18,22,23]. Most of our heat flow values are lower than ~45 mW $m^{-2}$, with only a few estimates higher than 70 mW $m^{-2}$ (Figure 4c). Several of these high values are found on, or close to, rifts structures, and are similar to those previously obtained for some rifts, coronae or even ribbon terrains[19-21]. Rifts and coronae correspond to (past or present) actively deforming lithosphere, with expected heat flows higher-than-average (see below).

Some works[16,17,19] obtained heat flows higher than our results, using high thermal conductivities of 3 or 4 W $m^{-1}$ $K^{-1}$ for the entire mechanical lithosphere. A study[16] of flexure induced by coronae loading obtained an average heat flow of ~95 mW $m^{-2}$ for a thermal conductivity of 4 W $m^{-1}$ $K^{-1}$. This value for the thermal conductivity is not realistic for the mantle lithosphere or the crust; because the thin mechanical thicknesses obtained by these authors must be mostly, or even entirely, within the crust, their average heat flow should be scaled to ~48 mW $m^{-2}$ for a thermal conductivity of 2 W $m^{-1}$ $K^{-1}$. Coronae could originate by energetic, although transitory, processes[53]. These processes could have had limited effect on the long-term strength of the lithosphere, and therefore local heat flows obtained for individual coronae are expected to be high, but not affecting significantly to the regional background heat flow (below we return on the potential effect of active plumes on the estimation of the total heat loss of Venus). On the other hand, an average heat flow of ~53 mW m$^{-2}$ was derived[17] for domical volcanoes not near coronae using a crustal thermal conductivity of 3 W m$^{-1}$ K$^{-1}$; scaling that value for our preferred crustal thermal conductivity we obtain ~35 mW m$^{-2}$, consistent with our results.

Smrekar et al.[19] presented a larger set of heat flow estimates from Coronae and Rifts structures. These authors found an average of 77.5 mW m$^{-2}$ for coronae. For the Parga and Dali Chasmata rift systems they found heat flows of, respectively, 68-104 mW m$^{-2}$ and 21-80 mW m$^{-2}$. However, their heat flow estimates are overestimated. On a hand, they used a thermal conductivity of 3 W m$^{-1}$ K$^{-1}$ for the entire mechanical lithosphere. On the other hand, much of their thermal gradients through the mechanical lithosphere are not consistent with their values for the temperature defining the base of the mechanical lithosphere, surface temperature and mechanical thickness (see their Supplementary Tables 1 and 3). For example, for Parga Chasma they quoted linear thermal gradients of 22.7-34.6 K km$^{-1}$, whereas that using their parameter values the thermal gradients should be 20.8-29.5 K km$^{-1}$, which, for a more realistic thermal conductivity of 2 W m$^{-1}$ K$^{-1}$, translate to heat flows between 42 and 59 mW m$^{-2}$; similarly, for Dali Chasma the heat flow would be 14-53 mW m$^{-2}$. Recalculating in the same way their heat flows for coronae an average of 46 mW m$^{-2}$ is obtained.

Smrekar et al.[19] also stated that the global elastic thickness map of Anderson and Smrekar[31] implies heat flows of 39-116 mW m$^{-2}$ for ~40% of the planet surface ($T_e$ values of 5-15 km) excluding tesserae, and ≤11-23 mW m$^{-2}$ for other ~45% of the surface ($T_e$ values of 25-55 or more km), again for a thermal conductivity of 3 W m$^{-1}$ K$^{-1}$. For the low $T_e$ range these heat flow values should be scaled to 26-77 mW m$^{-2}$ for a thermal conductivity of 2 W m$^{-1}$ K$^{-1}$. For the high $T_e$ range there could also have a significant contribution of the mantle thermal conductivity (3 W m$^{-1}$ K$^{-1}$; see Methods): considering a same weight for crustal ad mantle thermal conductivities, the heat flow should be scaled to ≤9-19 mW m$^{-2}$. The Anderson and Smrekar[31] global map does not have plate curvatures and therefore whichever heat flow derived from it is an upper limit (as in our own model). Thus, there is no support for widespread terrestrial-like heat flows on Venus.

Our regional background heat flows are consistent, or at least compatible, with the observed geology. For example, we obtain low heat flows at the central part of the BAT anomaly region, which is bordered by rift systems and plume-related features[46]. The presence of active rifting or mantle plumes does not necessarily imply high background heat flows. Indeed, Earth continental rifts can develop in areas of comparatively low background heat flows[54]: for example, the African Rift is crossing cratons characterized by a strong lithosphere and high $T_e$ (Ref. 55). The heat flow in rifts themselves may be high, and we find heat flows of even 66 mW m$^{-2}$ on Parga Chasma, 71 mW m$^{-2}$ on Devana Chasma, and 178 mW m$^{-2}$ on Dali Chasma (Figure 4b). In any case, the heat loss from continental rifts (i.e., non-mid-ocean spreading rifts) is lower than 1% of the total in continental areas[2], and therefore their contribution to the global heat budget is not significant. Otherwise, terrestrial oceanic rifting associated to mid-ocean spreading greatly contributes to the heat loss of the Earth, but there is nothing on Venus, to our knowledge, comparable a mid-ocean spreading. Similarly, plume activity on the Earth can pierce the lithosphere of cold cratons or old oceanic areas[2]. The total heat loss from active terrestrial plumes is 2-4 TW (Refs. 2,56), and although significant it is only around 5-10% of the total; a roughly similar plume contribution to the total heat loss, not recorded in the global lithosphere strength, could be considered for Venus (see below).

There are an enormous number of volcanoes on the Venusian plains[57], but the relation between the presence of volcanic edifices and high background heat flows is not straightforward. However, much of our highest heat flows are obtained for the plains north of Aphrodite Terra, which is consistent with the presence of a larger density of volcanic edifices[57], whereas the density of volcanoes is lower in the plains with lower heat flow in our model. Moreover, as above indicated, the average heat flow derived from the loading by volcanic domes[17] is similar to our nominal average heat flow.

Some geographical differences in heat flow may be related to age variations between regions. Because of the relatively young surface of Venus (~300-1000 Ma on average[28]), these age differences should mostly not be higher than a few hundred Ma. This time is comparable to the thermal time of the mechanical lithosphere, and therefore our map is informative of the last hundreds of millions of years of the planet. The thermal time is a rough estimate of the time for thermal perturbations to be dissipated out through heat diffusion: taking a standard heat diffusivity value for rocks of $\sim 10^{-6}$ m$^2$ s$^{-1}$, the thermal time would be $\sim 10^2$ Ma for a mechanical lithosphere ~50-100 km thick. The mechanical lithosphere is the real layer supporting geological stresses and is usually thicker than the effective elastic thickness; for a thick and unbent lithosphere, the mechanical and effective elastic thicknesses are similar[35,58] (see Methods).

The effective elastic thickness is representative of the time when the topography and gravity signals were established[59]. Therefore, there cannot be subsequent lithosphere heating, since it would imply lithosphere thinning and $T_e$ reduction. Later cooling and lithosphere thickening cannot be detected through $T_e$ analysis. Whichever potential process that can lower $T_e$ in a non-thermal way (e.g., stress relaxation) would result in an overestimation of the calculated heat flow. Thus, the implications of our results for the global dynamics of Venus are robust: both the calculated heat flow and their geographical variations are upper limits. Tessera terrains could be significantly older than the average surface, but since our calculations are upper limits our conclusions for recent times do not change.

Our model considers a basaltic crust for Venus. A felsic component in crustal plateaus should not significantly affect our conclusions. On one hand, the thermal conductivity of both felsic and basic crustal rocks at several hundreds of Celsius degrees is around 2 W m$^{-1}$ K$^{-1}$ (see Methods), and therefore the effect of a felsic component on the average thermal conductivity of the Venusian crust could be not significant; on the other hand, a felsic component would imply a proportionally weaker crust and a lower heat flow for a given $T_e$.

Fig. 7a shows the inverse of the bulk Urey ratio as a function of the total radioactive heat production of Venus; the inverse of the Urey ratio offers a more visual perspective of the amount of internal cooling/heating of a planet. The nominal total heat loss of Venus is set to 14.4 TW, but we also show the case for a heat loss of 17.2 TW (this value which can be taken as a very generous upper limit for the total heat loss of Venus if we consider global strain rates of $10^{-15}$ $s^{-1}$ and an additional ~10% heat loss from a hypothetical contribution from mantle plumes see [Methods]; this value is almost ~20% higher than our nominal result). The total radioactive heat production is calculated for the range deduced from terrestrial-based compositional scaling. Higher and lower total radioactive heat production values correspond to, respectively, chondritic and subchondritic compositions (see Methods). Because our heat flow map gives upper limits, the calculated inverse Urey ratios give an upper limit to the proportion between heat loss and radiogenic heat production and, therefore, to the amount of interior cooling (or conversely, a lower limit to the interior heating). For a total heat production of 14.4-17 TW or higher, the bulk Venusian interior would necessarily be heating up at present. The possibility of a sub-chondritic composition for the terrestrial planets, maybe related to early heavy erasing of their crusts by giant impacts, would be consistent with isotopic and mass balance evidence[60], and with the secular evolution of the temperature of the terrestrial mantle[61]. Otherwise, a recent comparison[62] of the dynamic topography of Baltis Vallis with models of stagnant-lid convection proposes a preferred average heat flow of 55-75 mW $m^{-2}$, but that work found unrealistically high crustal thicknesses. Moreover, the models used an internal heating rate of $5.2 \times 10^{-12}$ W $kg^{-1}$, higher than the values between 3.4 and 4.9 $\times 10^{-12}$ W $kg^{-1}$ deduced from compositional models[47-50,61]. This could explain the discrepancy with our results, especially considering the possibility of sub-chondritic heat-producing element abundances.

The used crustal radioactive heat production of 0.45 $\mu$W $m^{-3}$ provides a reasonable upper limit for Venus, but the actual value is not well known. Fig. 7a also shows the case without crustal heat sources. In this case, the interior is always heating up. Fig. 7b shows the total heat loss of Venus, the crustal contribution, and the mantle heat loss (given by the difference between both) for crustal heat production rates between 0 and 0.45 $\mu$W $m^{-3}$ and our nominal crustal thickness and strain rate model (and not considering any hypothetical heat loss from mantle plumes). We observe that the variation of the mantle heat loss is relatively small, between 9.2 and 11.0 TW, for the range of heat production considered. Thus, the nominal value of crustal heat production used here should not substantially influence the implications of our work for the dynamics of the Venusian mantle.

Large-scale geological structures are very different on Earth and Venus despite similar size and bulk properties. Earth shows tectonic structures mostly associated with plate boundaries[54], while the global tectonic pattern of Venus is characterized by distributed deformation and much more limited lithosphere subduction or spreading[19,63]. Our results show that from a thermal perspective, Venus is operating in its unique mode, which is globally less efficient in transmitting heat through the lithosphere than present-day terrestrial plate tectonics. Plate tectonics is a self-sustained process that efficiently cools the mantle and reduces the viscosity contrast across the silicate layer, which in turn favors a coupled lithosphere-convective interior[60]. Without global plate tectonics, interior cooling is less efficient, thus lowering lithosphere-convective coupling. In these conditions, other processes of heat transfer must be important.

Our heat flow model offers two important thermal constraints for understanding the Venusian dynamics, and successful interior evolution models should be consistent with them. Our results show that the geographical pattern of heat flow on Venus is less structured than on Earth and that, during the last hundreds of million years, the heat loss of Venus was roughly comparable to its radioactive heat production and that the interior cooling was, at most, limited.

## METHODS

### Effective elastic thickness ($T_e$) map

We use an updated version of the effective elastic thickness ($T_e$) model of Jiménez-Díaz et al.[29], derived from gravity and topography data and averaged onto 2º × 2º grids. The topography and gravity data were obtained from the spherical harmonic models SHTJV360u (ref. 64) and SHGJ180u (ref. 65) respectively. Both spherical harmonic models were truncated to degree and order 180.

The inversion for $T_e$ by Jiménez-Díaz et al.[29] was performed only on observed coherences with wavelengths >211 km. This corresponds to a flexural wavelength such that the minimum resolvable $T_e$ is ~14 km (see ref. 29 for details), which, therefore, imposes an artificial upper limit on the calculated heat flow. Therefore, here we carry out the inversion for the complete range of available and resolvable wavelet scales (equivalent Fourier wavelengths). The largest scale was chosen such that the longest equivalent Fourier wavelength was 6000 km (the side length of a tile); the smallest chosen scale corresponds approximately to the Nyquist wavelength of the gridded data (40 km; see ref. 29), which, if taken as a flexural wavelength, implies a minimum resolvable $T_e$ of ~2 km. After inversion, $T_e$ data at the edges of each tile (ten percent of a side length) were removed to mitigate possible remnant edge effects near the grid boundaries. As a final step, $T_e$ results were back-projected onto a geographic 2° × 2° grid and merged and gridded to produce a global map that combines the information from all tiles. The uncertainty in the determination of $T_e$ is taken as the 95% accuracy on the best-fitting value[29] (Fig. 2b). The spectral method of $T_e$ estimation does not allow for parameter uncertainties to be used to calculate uncertainties in the $T_e$ mapping; so the uncertainty map in Fig. 2b does not take these into account.

The effective elastic thickness model was obtained from the inversion of the Bouguer coherence function using the fan wavelet transform[30,66], modeled with a simple thin elastic plate subject to both surface and subsurface loads, following the load deconvolution procedure of Forsyth[67]. The wavelet method provides a coherence estimate at every point on the data grid. The analysis was performed in the Cartesian domain, dividing the surface of Venus into 36 overlapping areas (or 'tiles'), and projecting the gravity and topography in each of them to a Cartesian frame using an oblique Mercator map projection. The planar wavelet analysis for coherence and the $T_e$ inversion were then carried out for each tile.

The SHGJ180u harmonic coefficients of the gravity field beyond degrees 60–70 present large errors. While one could band-limit the gravity and topography data (excluding coefficients beyond degree 60 for instance) and perform the $T_e$ estimation again, the results might not be instructive, because that truncation would imply a minimum resolvable $T_e$ which would be highly artificial. Indeed, if the radius of Venus is 6052 km, then degree 60 is about 634 km wavelength. Using Simons and Olhede's (ref. 68) formula for the coherence transition wavelength gives 69 km as the minimum resolvable $T_e$ using the square real coherency (SRC) or coherence. So, if two clean gravity and topography models are used, both expanded out to degree and order 60, $T_e$ values less than 69 km cannot be obtained; if lower $T_e$ values are obtained, they would be artefacts of the inversion algorithm. Our choice to include higher harmonic degrees in the inversion does imply that some caution is required in the interpretation of the results, due to errors in the data at such degrees (see below).

## Errors in the gravity data and coherence versus admittance

When expanded, the error degree variances of the Venusian gravity field model will have regions where the errors are larger than the signal. Unfortunately, the $T_e$-estimation method does not provide for a direct propagation of data errors into the $T_e$ estimate. It does, however, include an error analysis based on (i) jackknifing of the coherence estimates, and (ii) estimates of the chi-squared misfit of the inversion (chi-squared misfits from the inversion, at all grid nodes were shown by Jimenez-Diaz et al.[29]). This approach will inherently betray noisy data regions. Furthermore, and importantly, the method inverts only the real component of the coherence, and it has been shown[30] that out-of-phase noise manifests predominantly in the imaginary component (which does not participate in the inversion, and hence our use of the SRC). So, although we cannot eliminate noise, we can decrease its influence while acknowledging its presence.

The coherence (or SRC as we use it) is characterised by a 'rollover' or transition wavelength, from high to low coherence. This rollover is quite steep, i.e. the transition is abrupt, occurring over a narrow range of wavelengths/frequencies. Furthermore, the coherence is very sensitive to variations in $T_e$, but relatively insensitive to other lithospheric parameters such as subsurface-to-surface loading ratio, and densities and depths of intra-lithospheric layers. In contrast, the admittance is much more sensitive to variations in all parameters, and importantly, it is sensitive over the entire spectrum rather than just a narrow band. So, while the admittance is less sensitive (but not immune) to noise than the coherence, the impact of noise is potentially more significant because it can alter the shape of the observed admittance function at 'random' frequencies in the spectrum, such that the inversion can easily misinterpret noise-induced variations as lithospheric-induced variations. Therefore, a $T_e$ analysis of grids generated from harmonic expansions is likely to be more robust when using the coherence than the admittance.

Accurate $T_e$ estimation with the coherence depends almost exclusively upon the relative positions of the actual rollover wavelength ($\lambda_r$) corresponding to the actual $T_e$, and the minimum wavelength ($\lambda_{min}$) in the harmonic model: if $\lambda_{min} > \lambda_r$ then $T_e$ cannot be reliably estimated; if $\lambda_{min} < \lambda_r$ then it can. In this regard, the harmonic degree of expansion makes no difference: if a harmonic model is noise-free up to degree 60 and we expand it to degree 180, then as long as the rollover wavelength falls in the noise-free bandwidths (i.e. $\lambda_r > 634$ km, the wavelength corresponding to degree 60 on Venus), the extra (noisy) coefficients from degrees 61-180 will not affect the inversion of the coherence. Thus, large errors at high degrees does not contaminate the entire spectrum if the coherence/SRC rollover lies in wavenumbers with low error degree variance, and then the $T_e$ estimates will be robust[29].

The admittance, however, is sensitive to lithospheric parameter variations over its whole bandwidth: if there is sufficient noise at shorter wavelengths to alter the shape of the admittance, then the inversion will fit a different curve to the admittance than it would in the noise-free case, and the difference could be manifested by different densities, depths and/or $T_e$. Therefore, an expansion to degrees higher than 60 could potentially affect $T_e$ from the admittance for any $T_e$ value. Thus, when using the admittance all harmonics are important; with the coherence only those around the rollover are important. The previous $T_e$ map of Anderson and Smrekar[31] used the admittance, and leakage between harmonics affects the admittance more than the coherence. Furthermore, they used a windowing[32] whose leakage properties are not ideal[33].

Thus, whereas recognizing the limitations raised from the gravity data, we assume our $T_e$ mapping as a representation of the lithosphere strength of Venus, and use that mapping to derive the heat flow model.

## Temperature profiles and heat flow calculation

We use the effective elastic thickness of the lithosphere ($T_e$) to constrain a lithospheric temperature profile and thus surface heat flow by using the equivalent strength envelope procedure[35,36]. The relation described by this procedure between lithosphere strength, strength envelope and heat flow is well established for Earth[58,69,70]. Flexural stresses can exceed the strength of rocks at both the top and bottom of the plate. At shallow depths, the brittle strength of rocks is determined by Byerlee's rule, whereas at deeper levels, rheology is described by power-law creep, which is temperature-dependent. The depth at which ductile strength reaches a given small value (and below which strength does not increase again) can be used to define a mechanical thickness, $T_m$. $T_e$ can then be related to $T_m$ by vertically integrating the lesser of yield and bending stresses for a given plate curvature, effectively calculating the bending moment[34]. The relation between $T_e$ and $T_m$ depends on plate curvature: if plate curvature is zero, then $T_e = T_m$, and this case gives the higher heat flow for a given $T_e$. Because the wavelet methodology does not permit a direct calculation of plate curvatures, we only present an upper limit calculation of heat flows from $T_e$, making the assumption $T_e = T_m$. For $T_m > T_e$, the surface heat flows would be somewhat lower than our estimates.

We calculated an elastic thickness, $T_e^*$ (we use an asterisk (*) to describe this calculated elastic thickness, as opposed to the $T_e$ estimated from the wavelet methodology) from the strength envelope predicted for a given heat flow. The surface heat flow is obtained through an iterative procedure when $T_e^* = T_e$. In the next we describe the procedure for calculating $T_e^*$.

The power law describing ductile strength is

$$(\sigma_1 - \sigma_3)_d = \left(\frac{\dot{e}}{A}\right)^{1/3} \exp\left(\frac{Q}{nRT}\right), \qquad (1)$$

where $\dot{e}$ is the strain rate, $A$, $Q$, and $n$ are empirically determined constants, $R$ (= 8.31446 J $mol^{-1}$ $K^{-1}$) is the gas constant, and $T$ is the absolute temperature. Here, we use a strain rate of $10^{-16}$ $s^{-1}$ in the calculations: this is a typical strain rate for active terrestrial plate interiors[44], and it is a reasonable upper limit for the recent tectonic activity of Venus[45]. As a sensitivity test, we also considered a strain rate of $10^{-15}$ $s^{-1}$, which is a value typical of terrestrial plate boundaries; geodetic studies[71] constrain the strain rate in the Great African Rift system to be $\leq 6 \times 10^{-16}$ $s^{-1}$. Thus, the use of $10^{-15}$ $s^{-1}$ overestimates the heat flow for most of Venusian regions (maybe for rifts too), and for this reason it is not appropriate for the construction of a global model, but it is useful to put a generous upper limit for the total heat loss of Venus. For creep parameters we use constants from a flow law for dry dunites[72]: $A = 28840$ $MPa^{-n}$ $s^{-1}$, $n = 3.6$ and $Q = 535$ kJ $mol^{-1}$. This flow law is appropriate for modeling the ductile strength of the mantle lithosphere. The ductile strength of the venusian crust is usually modeled using experimental flow laws for dry diabases[21,24,73], but since dry diabases are only slightly weaker than dry dunites, and that we are performing a calculation of heat flow upper limits, we use the dry dunite rheology for the entire lithosphere. We assume 10 MPa as the ductile strength value defining the base of $T_m$ (refs. 36, 74). Some works have used 50 MPa for defining the base of $T_m$ (e.g., 13,15-18), but while the use of an exact value does not alter much the results[35], lower strength implies a higher temperature in the $T_m$ base, and hence higher heat flows, which is useful for calculating heat flow upper limits. Indeed, considering a given heat flow, lowering the "defining strength" value implies a higher deep for the base of $T_m$ and therefore a higher bending moment of the strength envelope. To compensate it, the heat flow must be increased with respect to the case with higher "defining strength".

Temperature profiles in the crust are calculated by assuming a homogeneous distribution of radioactive heat sources, and therefore the temperature at a given depth $z$ is given by

$$T_{z,c} = T_s + \frac{F_s z}{k_c} - \frac{H_c z^2}{2k_c} \ , \tag{2}$$

where $T_s$ is the surface temperature, $F_s$ is the surface heat flow, $k_c$ is the thermal conductivity of the crust, and $H_c$ is the volumetric heating rate of the crust. We use a surface temperature of 740 K, a value of 0.45 μW $m^{-3}$ for the heat production of the crust (see next section), and a thermal conductivity of 2 W $m^{-1}$ $K^{-1}$, a standard value appropriate for basaltic rocks[75]. Furthermore, the thermal conductivity of felsic and basic crustal rocks at several hundreds of Celsius degrees is around 2 W $m^{-1}$ $K^{-1}$ (refs. 41,42). For this reason, this value is useful here, even considering a felsic component for the Venusian crust. We calculate temperature profiles in the mantle lithosphere using

$$T_{z,m} = T_{cb} + \frac{F_m (z - T_c)}{k_m} \ , \tag{3}$$

where $T_{cb}$ is the temperature at the base of the crust, $F_m = F_s - H_c T_c$ is the mantle heat flow, $T_c$ is the thickness of the crust, and $k_m$ is the thermal conductivity of the lithospheric mantle. We use $k_m$ = 3 W $m^{-1}$ $K^{-1}$, a value appropriate for mantle rocks at temperatures of at least several hundred degrees Celsius[41-43], as expected at the upper mantle of Venus.

This forward procedure allows us to calculate $T_e^*$ from $T_c$ and $F_s$. We can then compare $T_e^*$ with observed $T_e$ to put constraints on $F_s$, considering uncertainties in each estimated value. We use a Monte Carlo method to sample $T_e$ within their uncertainty bounds and minimize the squared difference between $T_e^*$ and $T_e$. $T_c$ and $T_e$ values are extracted from the models described in the previous section and averaged onto 2° × 2° grids. For each grid pixel, we calculate the uncertainty of $F_s$ due to the uncertainty in $T_e$ determination. As indicated above, the other parameter values have been selected to maximize the obtained heat flows. The standard deviation of the $F_s$ results due to the uncertainty in $T_e$ determination (see previous section) is shown in Fig. 3b.

## Crustal radioactive heat production

Considering the presence of radioactive heat sources in the crust increases the obtained heat flow (again useful for an upper limit calculation) compared to the case without crustal heat production (see Equation 2). The radioactive heat production of the Venusian crust, as well as its vertical or regional variations, is poorly known. There are five surface measurements of heat-producing elements (HPEs) abundances from Venera 8, 9 and 10, and Vega 1 and 2 landers[37,40]. If the discrepant measurements by Venera 8 are discarded, then mean abundances for Th, U and K are, respectively, 2 ppm, 0.6 ppm, and 0.4 percent. We calculated heat production rates using the decay constants of Van Schmus[76], obtaining heat production rates of 0.36-0.44 μW $m^{-3}$ for the age range between the present-day and 1000 Ma. We, therefore, use an upper limit value of 0.45 μW $m^{-3}$ in our heat flow calculations.

## Total radioactive heat production

We calculate a range for the present-day total radioactive heat production of Venus from mass-scaling of terrestrial values. The uncompressed densities of Venus and Earth are similar, so Earth can be considered a good model for Venus. The terrestrial total heat production is not well constrained, and the proposed values rely on the compositional model. We here consider four possible compositional models: the CI chondrite-constrained model of McDonough and Sun[47], the pyrolite model of Lyubetskaya and Korenaga[48], the Enstatite chondrite-based model of Javoy and Kaminski[50], and the model of Jackson and Jellinek[49] based on the geochemistry of Nd and Sm. The respective total heat production of these models ranges from 19.8 to 13.7 TW (see ref. 61), values that we scaled directly to the mass of Venus.

## Crustal thickness ($T_c$) map

We use the crustal thickness ($T_c$) model of Jiménez-Díaz et al.[29], averaged onto a 2º × 2º grid. The model was obtained following the potential theory procedure[77], and was constrained assuming (1) observed gravitational anomalies arising only from relief along the surface and crust-mantle interface, and (2) constant crustal and mantle densities of, respectively, 2900 and 3300 kg $m^{-3}$. The crustal thickness map was obtained by assuming a mean crustal thickness of 25 km, and subtracting the relief along the crust-mantle interface from the surface topography.

The mean crustal thickness of Venus is not well known. The value of 25 km assumed by Jiménez-Díaz et al.[29] is consistent with numerical models of rifting in Venus[78], but somewhat higher than the mean value of around 20 km proposed by Maia and Wieczorek[18] from Airy isostasy of crustal plateaus. In any case, the heat loss would change little for the mean crustal thicknesses of 20 or 25 km. This is consequence of a trade-off between the opposite effects that the crustal thermal conductivity and heat production have on heat flow calculation. For thicker crusts, the higher contribution of the low crustal thermal conductivity decreases the obtained heat flow; otherwise, for thicker crusts the higher total crustal heat production increases the obtained heat flow. For thinner crusts the opposite is true. When $T_e < T_c$ the obtained heat flow does not change (there is not contribution from the lithosphere mantle), and therefore the exact thickness of the crust does not affect the higher heat flow results, which are derived from low $T_e$ values. For $T_e$ = 40 km (the mean value for Venus), the heat flow of our nominal model would be 0.3 mW $m^{-2}$ higher for $T_c$ = 20 km than for $T_c$ = 25 km. Thus, considering 20 or 25 km for the mean crust thickness, the total heat loss only would change by around 0.1 TW, which has been taken into account in our Results and Discussion sections.

**Total heat loss**

We calculate the total surface heat loss implied by our heat flow map by adding the contribution of all the 2º × 2º pixels in the grid. The contribution of each pixel is given by multiplying heat flow by pixel area (which depends on latitude). This gives the heat loss expressed in power units[1,2].

We also propose a generous upper limit for the total heat loss by considering a global strain rate of $10^{-15}$ $s^{-1}$ (which is not realistic for most regions; see above). We add an additional 10% to the total heat loss to account for a hypothetical contribution from active plumes ("hot-spots"). We select this value because the contribution from active plumes to the heat loss of the Earth is ~5-10% of the total[2,56], and here we use the high bound for analogy.

Finally, we calculate the total mantle heat loss by subtracting the crustal contribution from the surface heat loss. An upper limit to the mantle heat loss is obtained if crustal heat sources are not considered. Also, we note that the average uncertainty in the determination of $F_s$ (0.8 mW $m^{-2}$, see Results) translates to an uncertainty of the total heat loss lesser than 0.4 TW.

**Basalt melting and basalt/eclogite transition**

To evaluate the possibility of large-scale basalt melting or crustal eclogitization, we compare crustal temperatures with the solidus and liquidus temperatures and the temperature for the basalt/eclogite transition. Basalt solidus and liquidus temperatures as a function of pressure are based on Hess and Head[79]. The relation between temperature and pressure for the basalt-eclogite transition has been calculated using data from Herzog et al.[80].

## ACKNOWLEDGEMENTS

This work was supported by funding from the Spanish Agencia Estatal de Investigación through the project PID2022-140686NB-I00 (MARVEN). This paper is dedicated to the memory of Blanqui L.R.

## DATA AVAILABILITY

The digital models for effective elastic thickness, crustal thickness, and heat flow are freely available from https://doi.org/10.5281/zenodo.13138657.

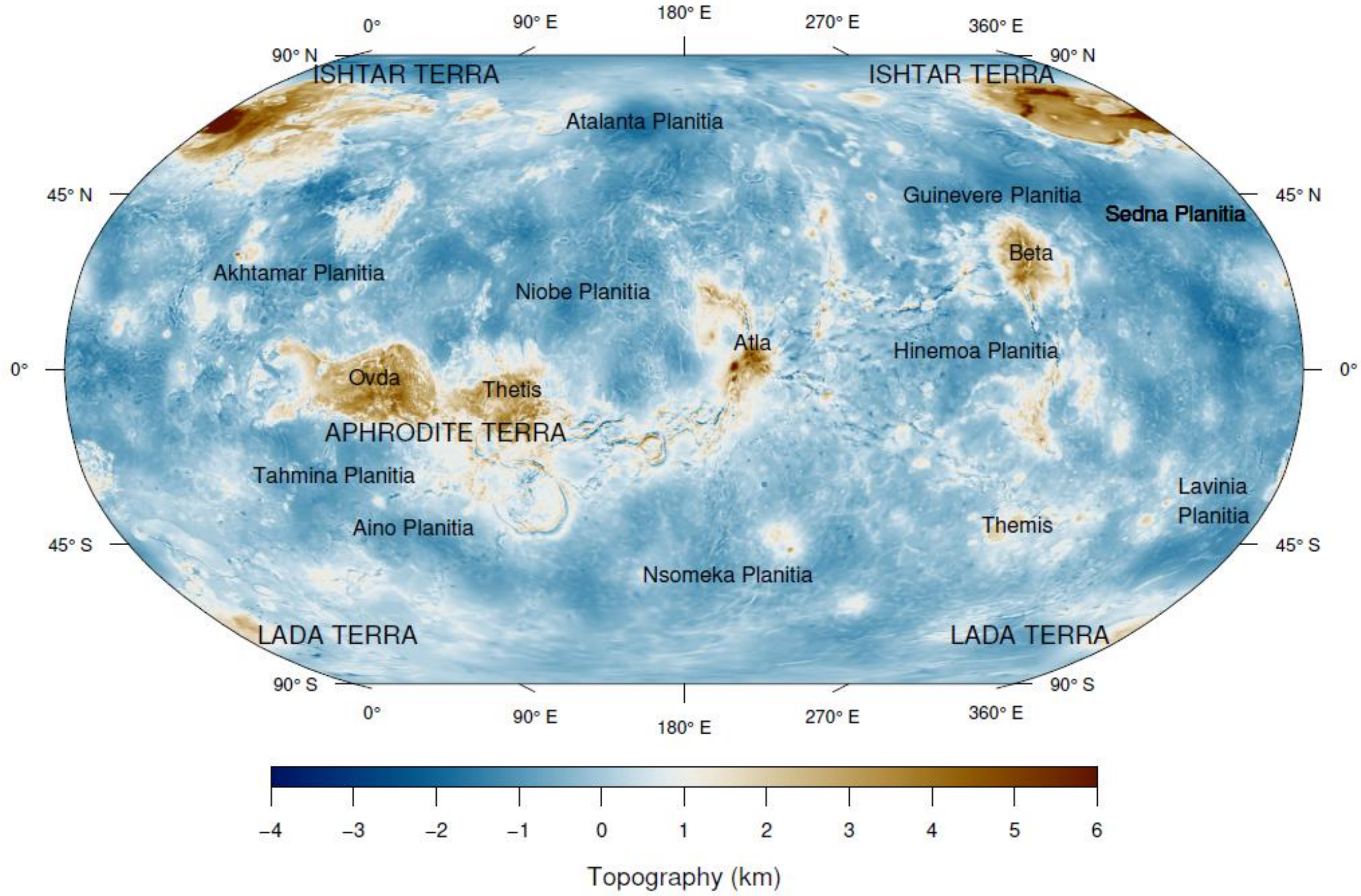

**Figure 1.** Topography map of Venus with names of the main regions for reference. The BAT anomaly region roughly comprises the terrains between Beta, Atla and Themis rises. All the global maps appear in a Robinson projection, with the central meridian corresponding to the 180ºE longitude.

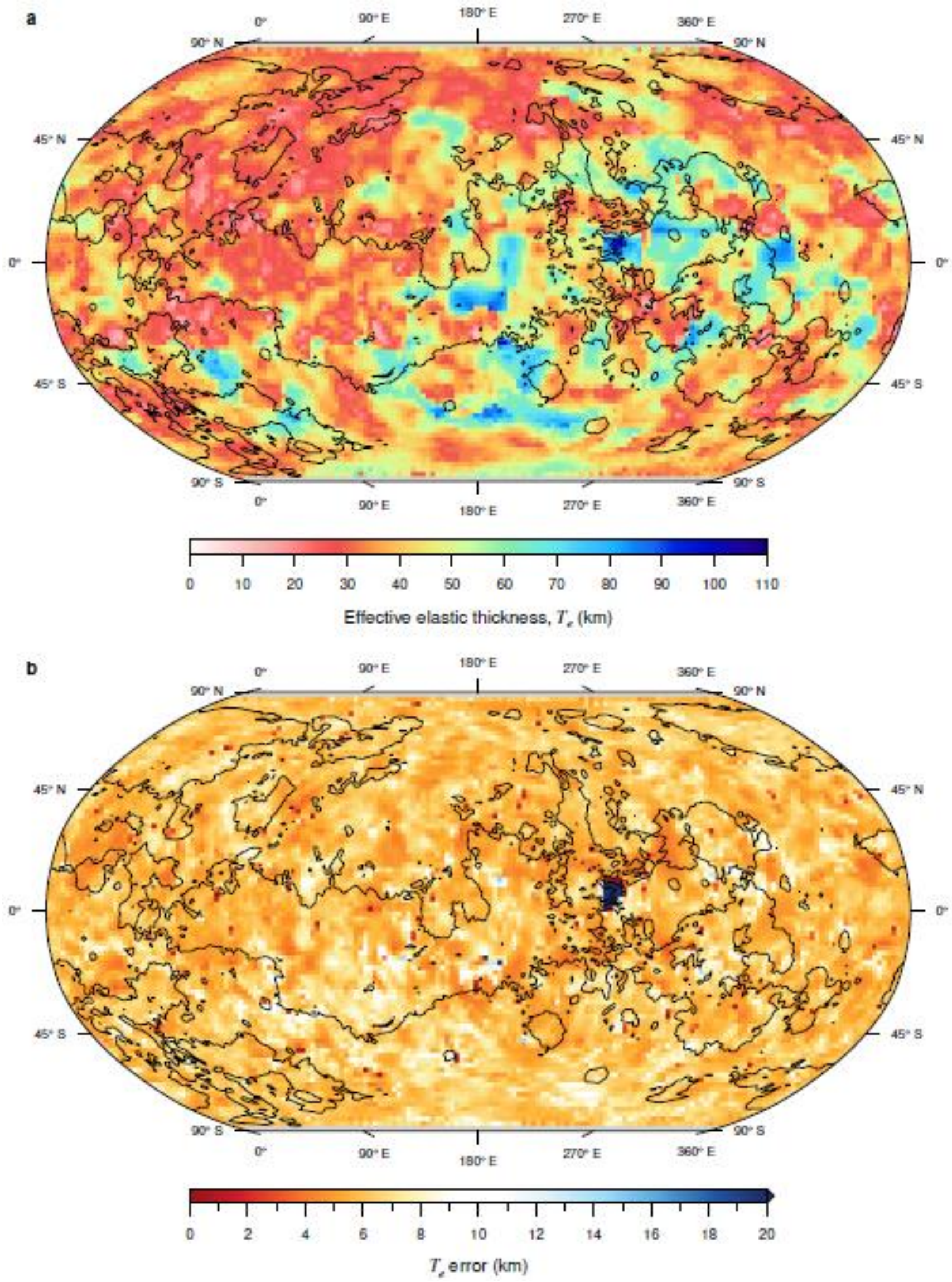

**Figure 2.** Effective elastic thickness model for Venus. (*a*) Global map of the effective elastic thickness ($T_e$) of Venus with a resolution of 2º × 2º; the total $T_e$ range is between 4.8 and 105 km. (*b*) Error in the determination of $T_e$ due to the incertainty in the wavelet analysis, given as 95% confidence limits on the best-fitting value.

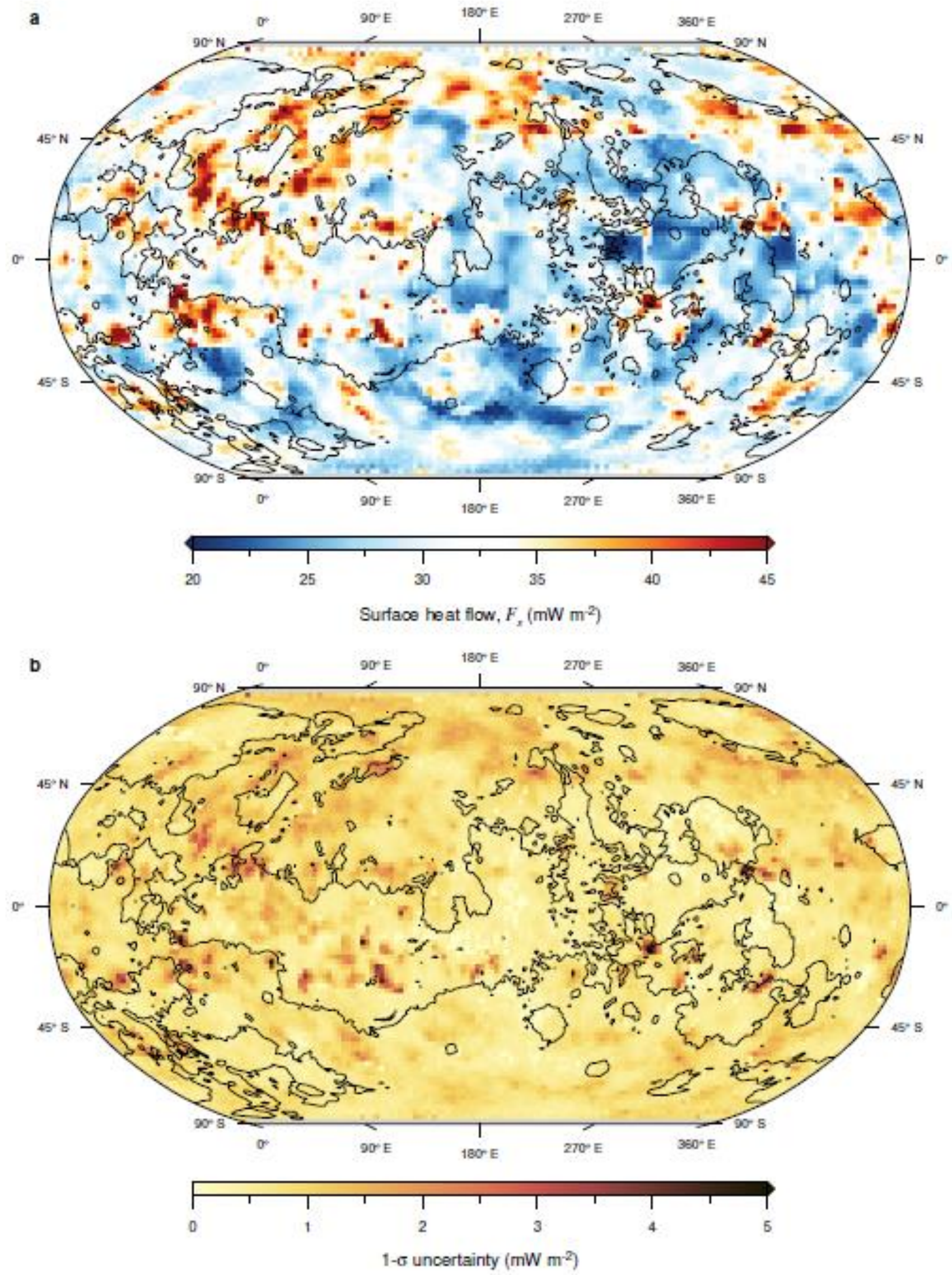

**Figure 3.** Global heat flow map for Venus. (*a*) Global map of the heat flow of Venus obtained from the effective elastic thickness of the lithosphere ($T_e$). Heat flows represented are the upper limit obtained for each grid pixel. The map resolution is 2º × 2º. The heat flow pattern is relatively smooth, with the ratio between maximum and minimum values lower than ten, although there are large-scale areas concentrating high and low heat flows. (*b*) Standard deviation of the calculated upper limit heat flows derived from the uncertainty in the determination of $T_e$ (and maintaining equal other parameters); due to the influence of crustal thickness in the calculation of heat flows, there is not a direct proportion in the respective uncertainties in $T_e$ (Figure 2b) and heat flow.

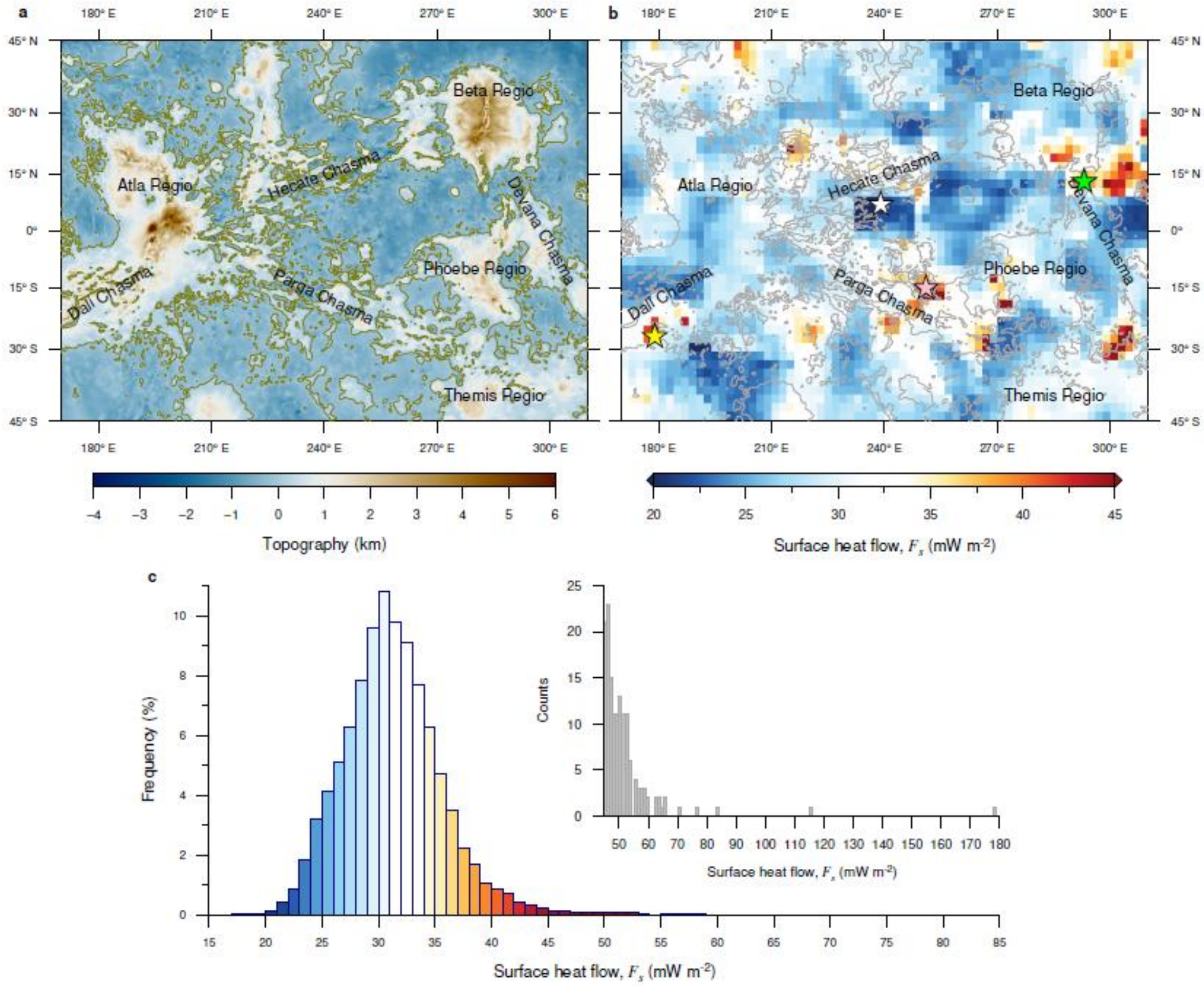

**Figure 4.** High heat flows on Venus. (*a*) Topography map of the BAT area showing the names of the main regions and associated rift systems; the central plains are the Hinemoa Planitia. (*b*) Same area but showing our heat model. Yellow star indicates the location of the highest heat flow on Dali Chasma (178 mW m$^{-2}$, the highest heat flow in our model); green star indicates the location of the highest heat flow (71 mW m$^{-2}$) on Devana Chasma; pink star indicates the location of the highest heat flow (66 mW m$^{-2}$) on Parga Chasma (66 mW m$^{-2}$); white star indicates the location of the lowest heat flow (17 mW m$^{-2}$) in our global model. (*c*) Histogram show the distribution of heat flows of our global model, including a detailed histogram for heat flows higher than 45 mW m$^{-2}$.

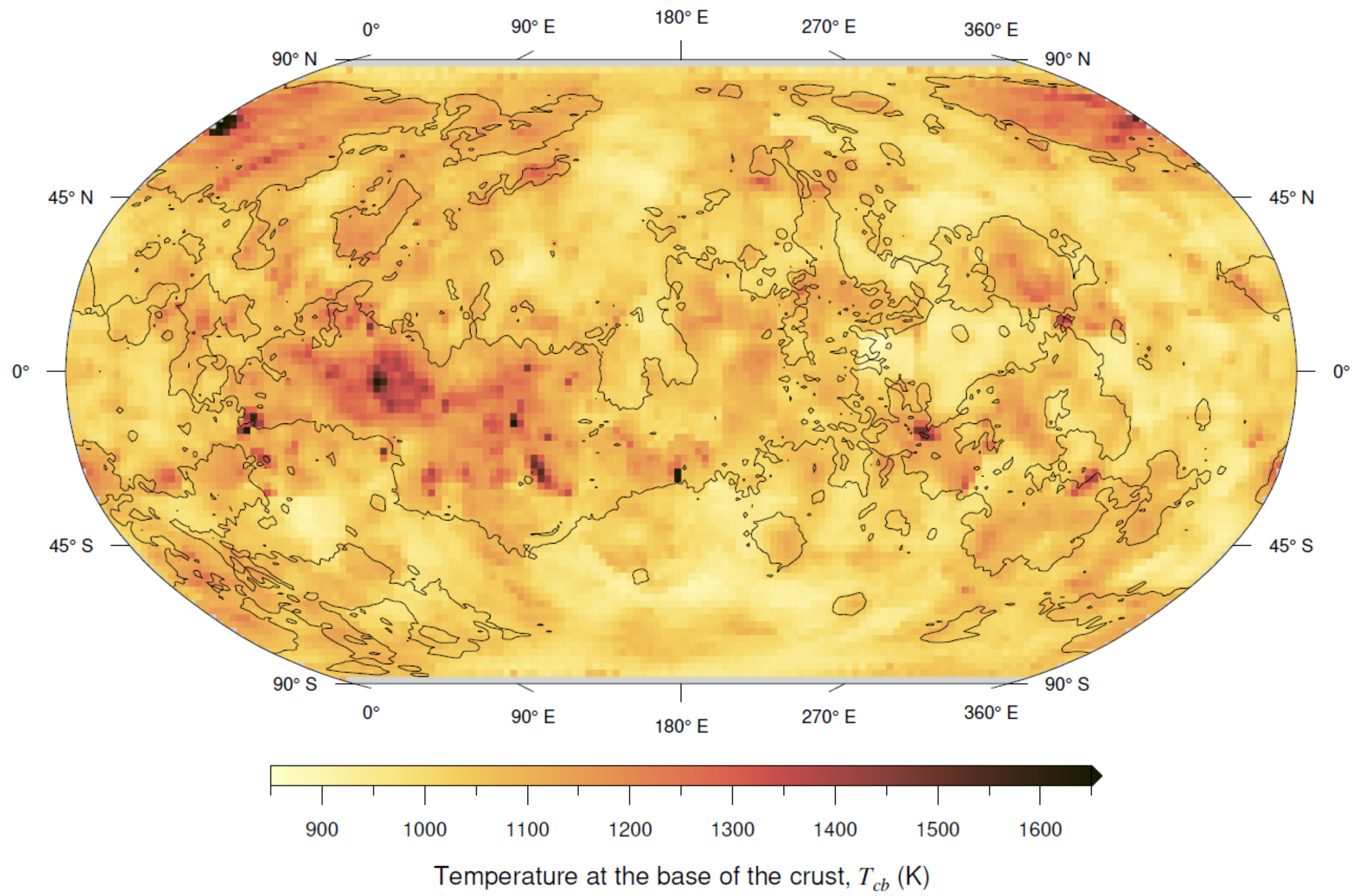

**Figure 5.** Global map of temperatures at the base of the crust obtained from our heat flow model.

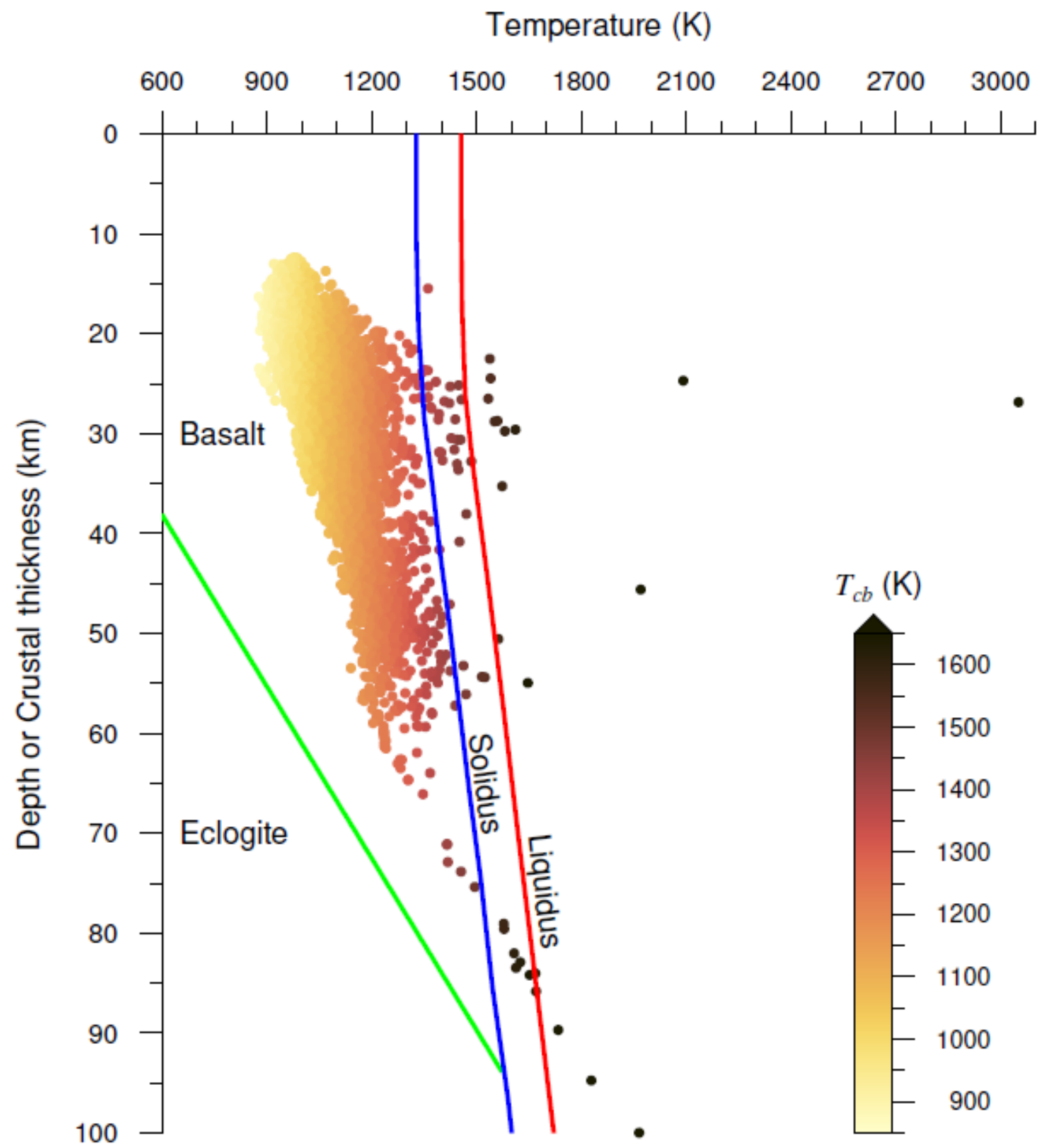

**Figure 6.** Crustal temperatures compared with conditions for basalt melting and eclogitization. Colored dots indicate temperatures at the base of the crust derived from our heat flow results (see Fig. 3a), compared with the conditions for solidus (melt initiation), liquidus (melt completion) and eclogitization of basalts (see Methods). Eclogitization is unexpected, and only a few temperatures at the base of the crust-crustal thickness points exceed solidus and liquidus conditions (59 and 17 points, respectively, on a total of 15692 points).

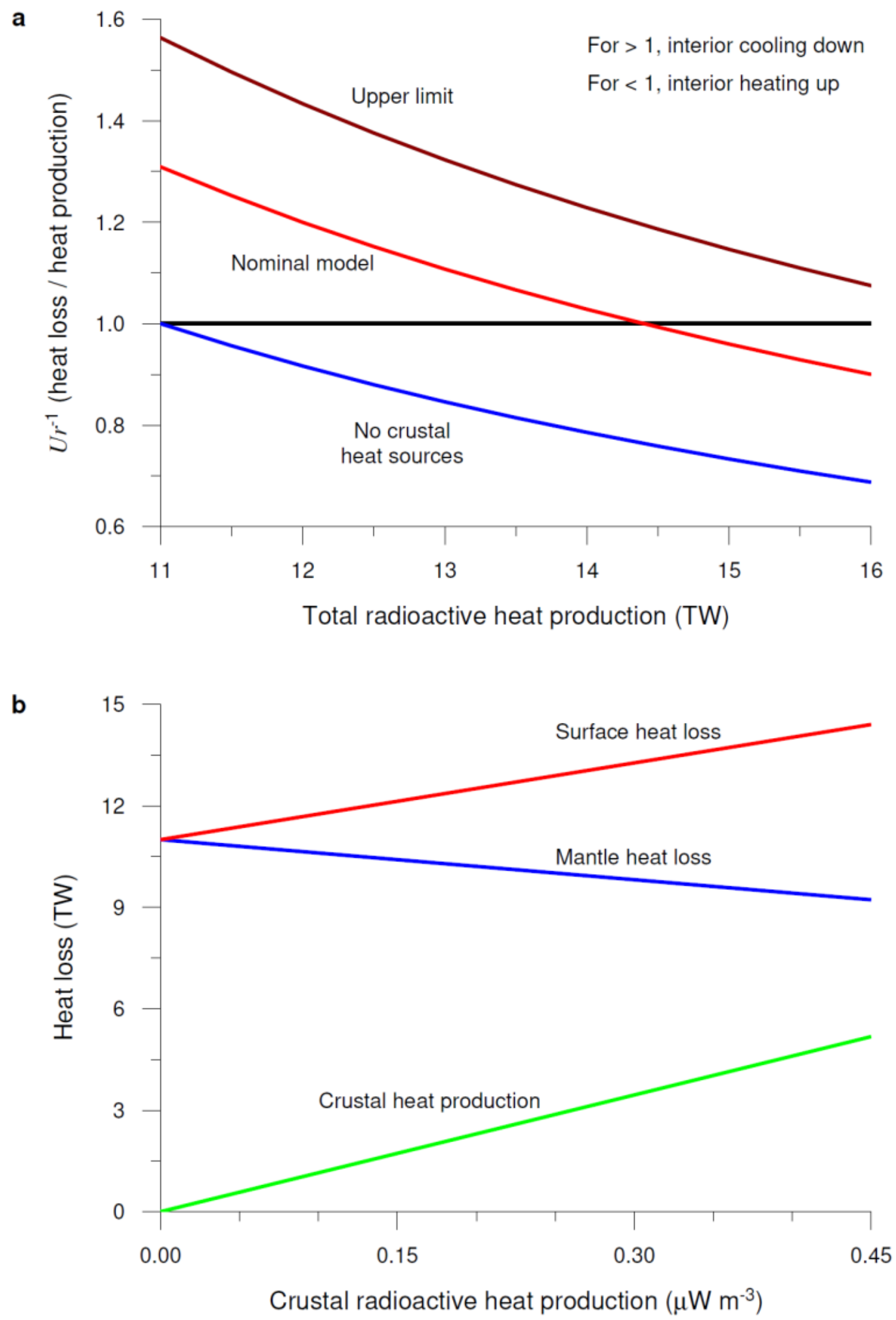

**Figure 7.** Total heat loss and heat budget of Venus. (*a*) The inverse of the Urey ratio (*Ur*) for bulk Venus. $Ur^{-1}$ gives the proportion between the total heat loss and the total heat production. It provides a direct representation of the heat budget of the planet and its implications for interior (considered as a whole) cooling or heating. $Ur^{-1}$ has been calculated from mass-scaling relations from a range of possible compositional models for Earth. The red curve represents our nominal heat loss of 14.4 TW, whereas the dark red curve represents the extreme case for a heat loss of 17.2 TW; both curves are calculated considering a radiogenic crustal heat production of 0.45 μW $m^{-3}$ (see Methods). For comparison, the blue curve represents the case without considering heat sources, which gives a total heat loss of 11.0 TW. The horizontal black line has been included to show the $Ur^{-1} = 1$ case. (*b*) Total surface and mantle heat loss in terms of the crustal heat production; for the highest heat production considered, 0.45 μW $m^{-3}$, the heat loss was set to our nominal value of 14.4 TW. The mantle heat loss is not very sensitive to crustal heat production.